\documentclass[10pt,conference]{IEEEtran}
\usepackage{cite}
\usepackage{amsmath,amssymb,amsfonts}
\usepackage{graphicx}
\usepackage{textcomp}
\usepackage{xcolor}
\usepackage[many]{tcolorbox}
\usepackage{booktabs}

\def\BibTeX{{\rm B\kern-.05em{\sc i\kern-.025em b}\kern-.08em
    T\kern-.1667em\lower.7ex\hbox{E}\kern-.125emX}}

\input{commands}

\newcommand{\HPCAck}{Part of the numerical simulations have been realized on the Linux HPC cluster Caliban of the High-Performance Computing Laboratory of the Department of Information Engineering, Computer Science and Mathematics (DISIM) at the University of L’Aquila.\xspace}

\newcommand{\EmeliotAck}{EMELIOT national research project, which has been funded by the MUR under the PRIN 2020 program (Contract 2020W3A5FY)\xspace}

\begin{document}

\title{On the Compression of Language Models for Code: An Empirical Study on CodeBERT}

\author{
\IEEEauthorblockN{Giordano d'Aloisio}
\IEEEauthorblockA{\textit{DISIM Department} \\
\textit{University of L'Aquila}\\
L'Aquila, Italy \\
giordano.daloisio@univaq.it}
\and
\IEEEauthorblockN{Luca Traini}
\IEEEauthorblockA{\textit{DISIM Department} \\
\textit{University of L'Aquila}\\
L'Aquila, Italy \\
luca.traini@univaq.it}
\and
\IEEEauthorblockN{Federica Sarro}
\IEEEauthorblockA{\textit{Department of Computer Science} \\
\textit{University College London}\\
London, UK \\
f.sarro@ucl.ac.uk}
\and
\IEEEauthorblockN{Antinisca {Di} Marco}
\IEEEauthorblockA{\textit{DISIM Department} \\
\textit{University of L'Aquila}\\
L'Aquila, Italy \\
antinisca.dimarco@univaq.it}
}

\maketitle

\begin{abstract}
Language models have proven successful across a wide range of software engineering tasks, but their significant computational costs often hinder their practical adoption. To address this challenge, researchers have begun applying various compression strategies to improve the efficiency of language models for code.
These strategies aim to optimize inference latency and memory usage, though often at the cost of reduced model effectiveness.
However, there is still a significant gap in understanding how these strategies influence the efficiency and effectiveness of language models for code.
Here, we empirically investigate the impact of three well-known compression strategies -- knowledge distillation, quantization, and pruning -- across three different classes of software engineering tasks: vulnerability detection, code summarization, and code search.
Our findings reveal that the impact of these strategies varies greatly depending on the task and the specific compression method employed. Practitioners and researchers can use these insights to make informed decisions when selecting the most appropriate compression strategy, balancing both efficiency and effectiveness based on their specific needs.
\end{abstract}


\section{Introduction}\label{sec:intro}
The growing adoption of transformer-based Language Models (LMs) has radically transformed the software engineering (SE)  field in recent years. These models have been effectively applied to a wide array of SE tasks, including vulnerability detection \cite{ding2024vulnerabilitydetectioncodelanguage}, code summarization \cite{Sun2024}, and code search \cite{Chen2024}, consistently demonstrating superior performance over traditional approaches \cite{hou2024largelanguagemodelssoftware}.
However, despite their impressive capabilities, the widespread adoption of these models is often hindered by practical challenges, particularly their high computational cost \cite{schwartz2019greenai}. The deployment of LMs typically requires computations across millions or even billions of learned parameters, resulting in significant memory demands and high inference latency.

To address this issue, AI researchers have developed various strategies over the past decade to reduce the size and computational cost of LMs, including techniques such as knowledge distillation \cite{Hinton2015}, quantization \cite{Zafrir2019}, and pruning \cite{Sanh2020}. 
These ``\emph{model compression}'' strategies can reduce the memory demand of LMs and/or speed up their inference times, albeit often at the cost of reduced effectiveness (i.e., prediction correctness).

These strategies have recently begun to gain attention in the field of software engineering \cite{Shi2023, Shi2024, Wei2023}.
For instance, Shi et al. \cite{Shi2023} applied knowledge distillation to drastically reduce the memory size of models of code.
Wei et al. \cite{Wei2023} have investigated the impact of quantization on code generation tasks regarding inference latency, memory consumption, and carbon footprint. 
However, despite these initial efforts, the broader impact of compression strategies on software engineering tasks remains largely unexplored.
Most existing research has focused on specific compression strategies applied to individual SE tasks, making it difficult to determine whether specific strategies perform better than others on particular tasks.
Additionally, it is unclear if the impact of different strategies varies by task or if they demonstrate similar behavior across different software engineering tasks.

In this study, we investigate the impact of different model compression strategies across three software engineering tasks: vulnerability detection (\emph{code classification}), code summarization (\emph{code-to-text generation}), and code search (\emph{text-to-code recommendation}). We fine-tune a well-known language model for code, CodeBERT \cite{feng2020codebert}, on each of these tasks. Subsequently, we assess how three model compression strategies -- namely, knowledge distillation, quantization, and pruning -- affect (i) the effectiveness of the LM in performing the task, (ii) inference latency, and (iii) the model's memory size.  Our results provide practitioners and researchers with guidelines on balancing the trade-offs between effectiveness and efficiency when selecting a model compression strategy.

In summary, the main contributions of this work  are:
\begin{itemize}[leftmargin=0.45cm]
    \item An extensive empirical evaluation of the impact of three compression strategies, namely knowledge distillation, quantization, and pruning on inference time, model size and effectiveness of LMs fine-tuned on the vulnerability detection, code summarization and code search tasks;
    %\item An extensive empirical evaluation of the impact of quantization in inference time, model size and effectiveness of LMs fine-tuned on the vulnerability detection, code summarization and code search tasks;
    %\item An extensive empirical evaluation of the impact of pruning in inference time, model size and effectiveness of LMs fine-tuned on the vulnerability detection, code summarization and code search tasks;
    \item Insights for practitioners and researchers on the adoption of those compression strategies;
    \item A publicly available replication package of our empirical study \cite{repl_package}.
\end{itemize}

\section{Background and Related Work}\label{sec:background}
\subsection{Language Models in Software Engineering Tasks}
Since its introduction by Vaswani et al. \cite{Vaswani2017}, the transformer architecture has become the de facto standard in language modeling. Transformer-based language models have achieved state-of-the-art performance across numerous natural language processing tasks \cite{radford2019language, raffel2023exploringlimitstransferlearning,devlin2019bertpretrainingdeepbidirectional}, and have recently gained traction in the software engineering field \cite{hou2024largelanguagemodelssoftware}.

This trend has led to the development of several LMs specialized in code, such as CodeBERT \cite{feng2020codebert}, CodeT5 \cite{wang-etal-2023-codet5}, and Codex \cite{chen2021evaluatinglargelanguagemodels}. 
These models are usually trained in a two-step process: (i) a pre-trained step using a self-supervised training objective aiming at providing the model with general knowledge about source code constructs and patterns, and (ii) a fine-tuning step aiming at tailoring the model for the software engineering task at hand.

Over the past few years, LMs for code have been fine-tuned to automate a wide range of SE tasks \cite{hou2024largelanguagemodelssoftware}. For example, CodeBERT -- an encoder-only LM based on the BERT architecture \cite{devlin2019bertpretrainingdeepbidirectional} -- has been widely and successfully applied to code summarization \cite{Gu2022}, vulnerability detection \cite{Zhou2021}, and code search \cite{feng2020codebert}, among others \cite{hou2024largelanguagemodelssoftware}.
 
\subsection{Compression Strategies for Language Models}

A major obstacle to the practical adoption of language models has always been their significant computational cost \cite{schwartz2019greenai, hou2024largelanguagemodelssoftware}.  To address this issue and increase their sustainability, the AI community has developed various strategies to reduce the memory demands and inference latency of these models. Nowadays, the three most popular model compression strategies are \cite{Shi2024,Shi2023}: \emph{knowledge distillation} \cite{Hinton2015}, \emph{quantization} \cite{Zafrir2019}, and \emph{pruning} \cite{Sanh2020}.

\subsubsection{Knowledge distillation} is a technique in which a smaller model (i.e., the ``\textit{student}'') is trained to replicate the behavior of a larger, pre-trained language model (i.e., the ``\textit{teacher}''). This compression strategy results in a model that demands less memory and provides faster inference time. However, since student models usually have thinner and shallower neural networks, they often struggle to fully capture the knowledge embedded in the larger models. This limitation typically leads to a reduction in the LM capabilities compared to the teacher model  \cite{sanh2020distilbertdistilledversionbert}. %For instance, DistilBERT, a popular distilled version of BERT, is approximately 60\% smaller and 40\% faster than the original, but it tends to perform worse on more complex tasks that require deeper language comprehension \cite{sanh2020distilbertdistilledversionbert}.

\subsubsection{Quantization} is a compression strategy that reduces the precision of the model's weights, converting them from full-precision (e.g., 32-bit floating point) to lower precision representations (e.g., 8-bit integer). Two broad categories of quantization strategies exist: \emph{post-training quantization} and \emph{quantization-aware training}. Post-training quantization generates a quantized model from an existing full-precision model without requiring additional training or fine-tuning. This strategy is a popular choice due to its low computational cost. However, it is more susceptible to quantization noise. In contrast, quantization-aware training involves training the model from scratch while incorporating simulated quantization operations to mitigate the quantization noise. Although this approach can produce a more effective model, its high training costs can make it often impractical.

\subsubsection{Pruning} aims to make the language model more efficient by removing neural network weights considered less critical for the model’s effectiveness \cite{LeCun1989}. Various pruning methodologies exist. For instance, \textit{structured pruning} modifies the model's architecture by removing entire structures within the neural network, such as neurons, filters, or even layers \cite{li2017pruning}. Conversely, \textit{unstructured pruning} targets individual weights \cite{guo2016dynamicnetworksurgeryefficient}, removing the less relevant ones (e.g., those close to zero). Pruning can be applied either to individual layers of the network (\textit{layer-wise pruning})\cite{molchanov2017pruningconvolutionalneuralnetworks}, or across the entire model (\textit{global pruning}) \cite{gale2019statesparsitydeepneural}.

\subsection{Compression Strategies in Software Engineering Tasks}
Model compression strategies have recently gained relevance in the software engineering literature. Shi et al. \cite{Shi2023} introduced \emph{Compressor}, a knowledge distillation-based approach, evaluated on CodeBERT \cite{feng2020codebert} and GraphCodeBERT \cite{guo2021graphcodebertpretrainingcoderepresentations} for two code classification tasks (i.e., vulnerability detection and clone detection). Their results demonstrate that \emph{Compressor} can considerably accelerate inference time with minimal impact on model effectiveness. In a subsequent study \cite{Shi2024}, the authors expanded their approach to tackle energy consumption and carbon footprint concerns.
Wei et al. \cite{Wei2023} conducted an empirical evaluation of quantized models on code generation tasks, examining resource usage, carbon footprint, accuracy, and robustness. They found that quantization, under specific settings, can substantially enhance model efficiency with negligible accuracy or robustness trade-offs.
Additionally, Sun et al. \cite{Sun2024} explored dynamic inference as a method to speed up code completion.
Despite these advancements, there remains a noticeable gap in comprehensive studies that systematically investigate the effects of different model compression strategies across various software engineering tasks.

With this study, we aim to fill this gap by performing an extensive empirical evaluation of the impact of the three most adopted compression strategies -- knowledge distillation, model quantization, and model pruning -- on the inference time, model size, and prediction's effectiveness of an LM fine-tuned for three widely adopted SE tasks -- vulnerability detection, code summarization, and code search.
%\textcolor{red}{quindi? come superiamo lo stato dell'arte? aggiungerei un paragrafo/box/sottosezione che dica come andiamo oltre. Forse è il primo paragrafo della sezione che segue...perchè non lo mettiamo qui come sottosezione? oppure come sezione a parte? lo evidenzierei dopo i related}

\section{Empirical Study Design}\label{sec:experiments}
The \emph{goal} of this study is to analyse the impact of compression strategies on the \textit{efficiency} (i.e., in terms of inference time and model size) and \textit{effectiveness} (i.e., in terms of prediction correctness) of language models for code.
Specifically, we investigate the impact of three compression strategies -- knowledge distillation, pruning, and quantization -- on CodeBERT models fine-tuned for three SE tasks: vulnerability detection, code summarization, and code search.
We selected these tasks due to their relevance in software engineering \cite{ding2024vulnerabilitydetectioncodelanguage,zhang2020software,Chen2024,Sun2024,wang2020trans} and because they span diverse categories, namely code classification (vulnerability detection), code-to-text generation (code summarization), and text-to-code recommendation (code search). We use CodeBERT as the reference language model due to its popularity in the software engineering literature \cite{hou2024largelanguagemodelssoftware} and its versatility in handling classification, generation, and recommendation tasks \cite{feng2020codebert}.

Our research is driven by the following research questions:

\begin{enumerate}
	\item[\textbf{RQ$_1$}] \emph{\rqone} 
	\item[\textbf{RQ$_2$}] \emph{\rqtwo} %Here, we examine the effects of different compression strategies on models fine-tuned for code summarization, focusing on both efficiency (inference time, model size) and effectiveness (the quality of generated summaries). %This RQ focuses on analysing the impact of compression strategies on an LLM fine-tuned on code summarization task (i.e., code-text generation).
	\item[\textbf{RQ$_3$}] \emph{\rqthree} %This RQ aims to assess how model compression strategies alter the efficiency and effectiveness of models in code search tasks. %\textcolor{red}{le 3 RQ differiscono sul task di SE, perchè abbiamo scelto questi task e non altri?le 3 RQ si ripetono troppo, forse si può sintetizzare e dire di più sullo specifico task}
%Specifically, it seeks to determine how these strategies affect inference time and model size, and to evaluate their impact on the relevance of code search results.  $$%This RQ focuses on analysing the impact of compression strategies on an LLM fine-tuned on code search task (i.e., text-code recommendation).
\end{enumerate}

%to investigate and compare how different model compression strategies affect efficiency metrics, such as the time required by the model to solve a task (i.e., inference time) and the model memory size, for each of the three tasks analysed herein. Additionally, we evaluate how these strategies influence the quality of the model output (i.e.,effectiveness) assessed by state-of-the-art metrics for the task at hand (see Section ). Last, but not least we do explore the trade-offs between efficiency and effectiveness for each of the tasks analysed herein.

%In the following, we present the SE tasks and the relative benchmark implementing them. Next, we describe the compression strategies and efficiency and effectiveness metrics employed in our evaluation.

%\input{tables/experiment}

%\subsection{Methodology}
%\label{sec:methodology}

\begin{figure}
    \centering
    \includegraphics[width=0.9\linewidth]{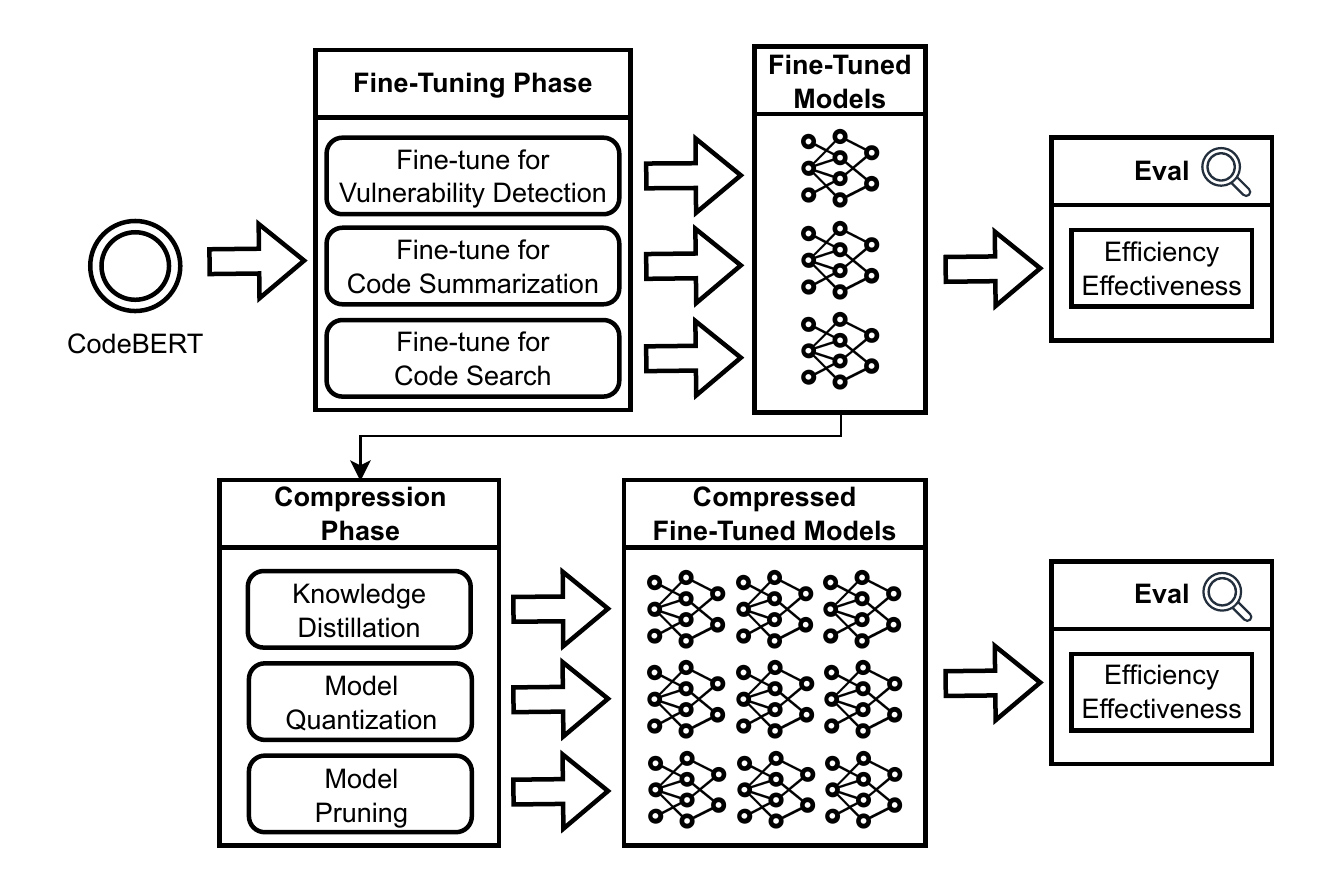}
    \caption{Experimental Methodology}
    \label{fig:experiment}
\end{figure}

To answer these RQs we carry out an empirical investigation based on the methodology described in the following. 
Figure \ref{fig:experiment} provides an overview of our experimental methodology. We first fine-tune CodeBERT for each SE task of interest and collect the corresponding effectiveness and efficiency metrics. Next, we apply each of the three compression strategies individually to produce compressed versions of the fine-tuned models. Finally,  we compare the efficiency and effectiveness metrics of the compressed models with those of the original fine-tuned CodeBERT model to assess the impact of each compression strategy.

\begin{table}[tb]
\centering
\caption{evaluation metrics}
\label{tab:metrics}
\resizebox{\columnwidth}{!}{%
\begin{tabular}{l|c|c|cc}
\toprule
& \textbf{Task} & \textbf{Task} & \multicolumn{2}{c}{\textbf{Metrics}} \\
 &  &  \textbf{Category} & \multicolumn{1}{c}{\textbf{Effectiveness}} & \textbf{Efficiency} \\ \midrule
 \midrule
\textbf{RQ$_1$} & \begin{tabular}[c]{@{}c@{}} Vulnerability\\ Detection\end{tabular} & \begin{tabular}[c]{@{}c@{}}Code-Code\\ Classification\end{tabular} & \multicolumn{1}{c|}{\begin{tabular}[c]{@{}c@{}}Accuracy\cite{rosenfield_coefficient_1986}\\ F1 Score\cite{taha_metrics_2015}\\ MCC\cite{chicco2020advantages}\end{tabular}} &  \\ \cmidrule(r){1-4}
\textbf{RQ$_2$} & \begin{tabular}[c]{@{}c@{}}Code\\ Summarization\end{tabular} & \begin{tabular}[c]{@{}c@{}}Code-Text\\ Generation\end{tabular} & \multicolumn{1}{c|}{\begin{tabular}[c]{@{}c@{}}Bleu\cite{papineni_bleu_2002}\\ BERTScore\cite{zhang2019bertscore}\\ SIDE\cite{mastropaolo_evaluating_2023}\end{tabular}} & \begin{tabular}[c]{@{}c@{}}Inf. Time (sec.)\\ Model Size (MB)\end{tabular} \\ \cmidrule(r){1-4}
\textbf{RQ$_3$} & \begin{tabular}[c]{@{}c@{}}Code\\ Search\end{tabular} & \begin{tabular}[c]{@{}c@{}}Text-Code\\ Search\end{tabular} & \multicolumn{1}{c|}{\begin{tabular}[c]{@{}c@{}}MRR\cite{mrr_score}\\ MRR@1\cite{mrr_score}\\ MRR@5\cite{mrr_score}\end{tabular}} &  \\ \bottomrule
\end{tabular}%
}
\end{table}

Table \ref{tab:metrics} shows the effectiveness and efficiency metrics used in our study. For effectiveness, we consider specific metrics depending on the software engineering task at hand. For instance, we use \textit{Accuracy}, \textit{F1 Score} and \textit{MCC} to measure the effectiveness of the LM for the vulnerability detection (i.e., code classification) task. For efficiency, we focus on two aspects: inference time (in seconds) and model memory size (in MB). We collect inference time metrics for both CPU and GPU environments, as compression strategies are frequently used to adapt language models for constrained hardware setups, such as desktop computers without dedicated GPUs. The experimental process took around ten days of machine execution over a CentOS HPC cluster equipped with 32 Intel(R) Xeon(R) Gold 6140M CPUs and two Nvidia A100 and A30 GPUs. 

%In the following subsections, we provide a detailed description of (i) the software engineering tasks studied and their corresponding experimental setups (Section \ref{sec:tasks}), (ii) the process used to generate the compressed models (Section \ref{sec:strategies}), and (iii) the evaluation metrics used to assess both efficiency (Section \ref{sec:effic_metrics}) and effectiveness (Section \ref{sec:effec_metrics}).

\subsection{Software Engineering Tasks}\label{sec:tasks}
\subsubsection{Vulnerability Detection}
In this task, the language model is prompted with a code function and asked to predict whether it contains a security vulnerability. The model produces a binary label indicating whether the code is vulnerable or not. We fine-tune and evaluate CodeBERT using the Devign dataset by Zhou et al. \cite{zhou2019devign}. This dataset contains 27,318 C functions extracted from two popular open-source projects (QEMU and FFmpeg). Each function is accompanied by a label that denotes whether the code contains a security vulnerability or not.

%\todo{if space allows for it, explain more details about this dataset to make our paper self-contained.}

\subsubsection{Code Summarization}
It is the task of automatically generating
natural language summaries (namely comments) for code
snippets. The language model is given a code function to produce code summary. We use a Sequence-to-Sequence (Seq2Seq) model, which includes CodeBERT as encoder layer and a six-layer Transformer as decoder layer. We fine-tune and evaluate CodeBERT on the CodeSearchNet dataset \cite{husain2019codesearchnet}. This dataset contains 2 million code-comment pairs extracted from open-source repositories written in different languages, such as Python, Javascript, Ruby, Go, Java, and PHP. For this task, we focus on the Java programming language for a total of 181,061 code-comment pairs.  

%\todo{if space allows for it, explain more details about this dataset to make our paper self-contained.}

\subsubsection{Code Search} Given a natural language sentence (i.e., comment), this task aims to retrieve semantically relevant code snippets. This is performed by first producing the embeddings of the comment and the code snippets. Then the semantically similar code snippets are ranked using the embedding similarity of sentence and code (through inner dot product). We fine-tune and evaluate CodeBERT using the CodeSearchNet dataset for the Python programming language, which contains a total of 280,634 code-comment pairs \cite{husain2019codesearchnet}. 
%\medskip

\begin{table}[tb]
    \centering
    \caption{datasets and lm hyper-parameters for each task}
    \label{tab:datasets}
    \resizebox{\linewidth}{!}{\begin{tabular}{l|c|c|c|c|c|c}
    \toprule
     & \textbf{Task} & \multicolumn{4}{c|}{\textbf{Dataset}} & \textbf{Hyper} \\
     & & \textbf{Name} & \textbf{Train} & \textbf{Val.} & \textbf{Test} & \textbf{Params.}\\
    \midrule
    \midrule
    \textbf{RQ$_1$} & \makecell{Vulnerability\\Detection} & Devign \cite{zhou2019devign} &   21,854 & 2,732 & 2,732 & \makecell{Epochs:  5\\Learning Rate: $2^{-5}$}\\
    \midrule
    \textbf{RQ$_2$} & \makecell{Code\\Summarization} & \makecell{CodeSearchNet\\(Java) \cite{husain2019codesearchnet}} & 164,923 & 5,183 & 10,955 & \makecell{Epochs:  10\\Learning Rate: $5^{-5}$}\\ 
    \midrule
    \textbf{RQ$_3$} & \makecell{Code\\Search} & \makecell{CodeSearchNet\\(Python) \cite{husain2019codesearchnet}} & 251,820 & 9,604 & 19,210 & \makecell{Epochs:  2\\Learning Rate: $2^{-5}$}\\
    \bottomrule
    \end{tabular}}
\end{table}

Following previous works \cite{Shi2023,Shi2024}, we reuse the pipeline provided by the CodeXGLUE benchmark \cite{DBLP:journals/corr/abs-2102-04664} for all the aforementioned tasks. CodeXGLUE is a popular benchmark which provides data and code for fine-tuning and evaluating LMs on different code-related tasks. We extended the pipeline by adding the code required to compress the LMs using knowledge distillation, quantization, and pruning. 

For evaluation, we use the train, validation, and test splits as well as the same model hyper-parameters provided by the CodeXGLUE benchmark. 
The size (in terms of number of items) of each dataset split and the model's hyper-parameters are reported in Table \ref{tab:datasets}.

\subsection{Compression Strategies}\label{sec:strategies}

%We analyse the impact of the three most adopted model compression strategies, i.e. \textit{Knowledge Distillation}, \textit{Model Quantization}, and \textit{Model Pruning}. 

\subsubsection{Knowledge Distillation}
%As said in Section \ref{sec:background}, Knowledge Distillation refers to the strategy of training a smaller model (called \textit{``student"}) by transferring the knowledge from the original fully-trained model (called \textit{``teacher"}). Unlike Quantization and Pruning,
Knowledge distillation is a computationally complex task, as it typically involves re-training the ``student'' model from scratch. Given the extensiveness of our experimental setup, we opted not to retrain a distilled model ourselves. Instead, we utilized a pre-trained distilled BERT model, namely DistilBERT \cite{sanh2020distilbert}, which we fine-tuned for the specific SE task of interest.
%For this reason, to improve the reliability of our evaluation, we employ DistilBERT \cite{sanh2020distilbert} as an implementation of Knowledge Distillation. DistilBERT is a distilled version of BERT, following the same architecture as the CodeBERT baseline model.
For vulnerability detection and code search tasks, we directly fine-tuned DistilBERT. For code summarization, we used DistilBERT as the encoder layer of a Seq2Seq model, which we then fine-tuned using the same procedure.

\subsubsection{Quantization}
Model Quantization reduces a model's size by changing its weights' precision from the standard \texttt{float32} to less precise data types. We apply \textit{post-training quantization} (see Section \ref{sec:background}) implemented by the Hugging Face's \textit{optimum-quanto} library.\footnote{\url{https://github.com/huggingface/optimum-quanto}} We chose this implementation because it requires minimum configuration and supports CPU and GPU. We tested three different applications of post-training quantization by reducing the weights to \texttt{int4}, \texttt{int8}, and \texttt{float8} (as, by the time we ran our experiments, the library allowed those three reductions). Following the library's documentation, we first quantized the fine-tuned model and then calibrated its activation functions using the validation set. Finally, following again the documentation, we \textit{frozen} the quantized weights before storing the model.

\subsubsection{Pruning}
%Model Pruning compresses a model by removing the less relevant parameters from the network. As described in Section \ref{sec:background}, there are different pruning strategies. 
In our experiments, we analyse the \textit{unstructured global pruning} implemented by the PyTorch library.\footnote{\url{https://pytorch.org/tutorials/intermediate/pruning_tutorial.html}} We have chosen this implementation because it does not change the internal structure of a network; hence, it does not require the re-training of the model after its application. Following the work of Gordon et al. on pruning BERT models \cite{gordon_compressing_2020}, for each task, we prune the weights of all the linear layers of the network using the \textit{L1 norm} as the selection strategy. The \textit{L1 norm} strategy uses the sum of the absolute values of a vector's components to determine the importance of structures within a neural network \cite{kumar2021pruning}. We analyse three different configurations of pruning: one in which we prune the 20\% of weights in the whole network (Prune 0.2), one in which we prune the 40\% (Prune 0.4), and one in which we prune the 60\% (Prune 0.6). As done for quantization, we first fine-tuned the models for each task and then pruned them. Finally, before storing the model, we removed the internal copy of the original weights created by the PyTorch library after the application of pruning.

%It is worth mentioning how, to increase the overall reliability of the evaluation, we adopted production-level implementation of the compression strategies released by companies working in the field.

%\subsection{Effectivenss Metrics}

%We evaluate the impact of compression strategies over two macro-dimensions, \textit{effectiveness} and \textit{efficiency}. Table \ref{tab:metrics} summarises the metrics employed on each RQ, which we describe in the following. All experiments have been executed on a CentOS HPC cluster equipped with Intel(R) Xeon(R) Gold 6140M CPUs and Nvidia A100 GPUs. 

\subsection{Efficiency Metrics}\label{sec:effic_metrics}

For efficiency, we indicate the performance of a model in terms of the time required to give a prediction (i.e., inference time) and the memory size of a model.

\subsubsection{Inference time} We measure the inference time for each batch of the testing set. Following the CodeXGLUE benchmark, we consider a batch size of 64 test instances for each task. The inference time is measured both on CPU and GPU (CUDA). For CPU, we use the \texttt{time} Python function, while for GPU, we use the \texttt{Event} class provided by PyTorch. Moreover, before computing the inference time on the GPU, we perform a series of warm-up iterations to avoid inconsistencies in the results. In addition, we apply GPU synchronization after each inference iteration.\\
To assess the impact of compression strategies, we compare the inference time measurements of each compressed model with those of the original fine-tuned CodeBERT model. To ensure rigor, we follow performance engineering best practices \cite{Traini2024, Jangali2023a, Zhang2023a, Laaber2020a}, specifically the approach proposed by Kalibera and Jones \cite{Kalibera2013a}, to build confidence intervals for the relative change in measurements statistics. Specifically, we construct the confidence interval for the median relative change in inference time using bootstrapping with 10,000 iterations, involving random resampling with replacement \cite{Kalibera2012a}. The main advantage of this technique, compared to others such as the Wilcoxon test \cite{woolson2005wilcoxon}, is that it provides a clear and rigorous account of the inference time change and the associated uncertainty \cite{Kalibera2012a,Kalibera2013a,Traini2021a}.
For example, this method allows us to state that a compressed model is faster than the original CodeBERT model by -30\%$\pm$2\% with 95\% confidence. We consider a difference to be statistically significant if the confidence interval is not greater than the percentage change.
%\todo{update with new statistic}
%In Section \ref{sec:results}, we report the median inference time and its standard variation among all batch iterations for a specific task.
%Moreover, we use the non-parametric Wilcoxon test \cite{woolson2005wilcoxon} to determine if there is a statistically significant difference between the inference time of the model with no compression strategy and the compressed models. We adopt this test because the experiment follows a \textit{one factor with multiple treatments paired comparison} design (i.e., all compression strategies have been evaluated on the same models) \cite{wohlin_experimentation_2012}. Following standard practices \cite{arcuri2011,wohlin_experimentation_2012}, we consider a statistical test significant if its p-value is $<0.05$. However, since we perform seven hypothesis tests for each RQ (i.e., one for each compression strategy), we apply the \textit{Bonferroni} correction \cite{weisstein2004bonferroni} and consider a test significant if its p-value is $< 0.05/7$.

\subsubsection{Model size}
To measure the model size, we save the model state in memory using the \texttt{save} function provided by PyTorch and then calculate its size using the \texttt{getsize} Python function. The value the function returns is converted to megabytes (MB). Although we performed this process on both CPU and GPU, as expected, the models’ size remained unchanged between the two environments. Therefore, we do not differentiate between CPU and GPU when reporting the models' size in Section \ref{sec:results}.

\subsection{Effectiveness Metrics}\label{sec:effec_metrics}
For effectiveness, we assess how good the predictions of a model are for a specific task. Given the heterogeneity of tasks involved in our evaluation, we used different metrics depending on the SE task being analyzed (see Table \ref{tab:metrics}).

\subsubsection{Vulnerability Detection} We use the \textit{Matthews Correlation Coefficient (MCC)}  as it considers all quadrants of the classification matrix (i.e., it gives a comprehensive overview of the model's performance). It has been shown to be a reliable measure when handling imbalanced data, as is often the case for vulnerability prediction \cite{zimmermann2010, JimenezFSE, MoussaDPmetrics, chicco2020advantages}. MCC is defined as a correlation factor between the true and predicted labels. It ranges from -1 to 1, where -1 means the model gives opposite predictions, 0 means random predictions, and 1 means perfect predictions.
In our study, we also report on \textit{F1 Score} \cite{taha_metrics_2015} and \textit{Accuracy} \cite{rosenfield_coefficient_1986} for compatibility with respect to previous work.
The F1 Score is defined as the harmonic mean between Precision and Recall. Accuracy is defined as the number of correct predictions over the whole predictions of a model. Both F1 Score and Accuracy values range from 0 to 1, where 1 is the best value. Although the use of Accuracy is deprecated for problems suffering from data imbalance \cite{MoussaDPmetrics}, we include this measure in our analysis for completeness as it is the metric employed by the CodeXGLUE benchmark. Still, we discourage its use in practice as done in previous work \cite{MoussaDPmetrics}. 

\subsubsection{Code Summarization} We employ three metrics that assess different aspects of the quality of a generated text \cite{sun_source_2024}. Bleu is the metric employed in the CodeXGLUE benchmark and is a standard metric adopted in natural language translation and, generally, text generation tasks \cite{papineni_bleu_2002}. It computes how similar a generated text is with respect to a reference baseline by comparing overlapping n-grams (contiguous sequences of n words) between the generated and reference texts. It is a metric of \emph{summary-summary text similarity}, but has been criticised for not considering the semantic similarity between two texts \cite{sun_source_2024,Wei2023}. For this reason, we extended the evaluation by including two additional metrics. BERTScore evaluates the quality of a generated text by comparing the similarity between the BERT embeddings of the generated text and the reference baseline \cite{zhang2019bertscore}. It is a metric of \emph{summary-summary semantic similarity}  \cite{sun_source_2024}. Finally, SIDE is a metric based on contrastive learning that measures how good a generated explanation is for a given code snippet without considering a reference baseline \cite{mastropaolo_evaluating_2023}. It is specific for code summarization tasks and is a metric of \emph{summary-code semantic similarity} \cite{sun_source_2024}. All metrics range between 0 and 1, where 1 is the optimum value.

\subsubsection{Code Search} We employ three versions of the Mean Reciprocal Rank (MRR) score \cite{mrr_score}. We adopt this metric because it is used in the CodeXGLUE benchmark and is by far the most commonly adopted metric in code search \cite{Xie2023}. MRR measures the average of the reciprocal ranks of the correct results for a set of code comments. The reciprocal rank for a single code comment is defined as the inverse of the rank position where the correct corresponding code appears in the list of retrieved results. In addition, we include two variations of MRR: MRR@1, which measures the proportion of code comments where the correct code appears in the first position, and MRR@5, which calculates the mean reciprocal rank based on the top five results. These metrics all range from 0 to 1, with 1 representing the optimal score.

\section{Empirical Study Results}\label{sec:results}
In this section, we report the result of our empirical evaluation. For each RQ, we discuss the impact of each compression strategy relative to the model's efficiency and effectiveness, as well as the trade-off between these two aspects.

\begin{table*}[ht!]
    \centering
    \caption{RQs 1-3: Efficiency and effectiveness of original and compressed code models for each of the SE tasks.}
    \label{tab:results}
    \begin{subtable}{\textwidth}
        \centering
        \resizebox{.8\textwidth}{!}{\begin{tabular}{l||c|c|c||c|c|c}
        \toprule
        \textbf{Compression} & \multicolumn{3}{c||}{\textbf{Efficiency}}  & \multicolumn{3}{c}{\textbf{Effectiveness}}\\
        \textbf{Method} & \textbf{CPU Inf. Time} & \textbf{GPU Inf. Time} & \textbf{Model Size}   & \textbf{Accuracy} & \textbf{F1} & \textbf{MCC} \\
        \midrule
        \midrule
        None & 15.902 (sec) & 0.011 (sec) & 499 (MB) & 0.630 & 0.541 & 0.247 \\
        \midrule
        \midrule
        Know. Distil. & -39.8\% $\pm$ 2.7\% & \textbf{-47.7\% $\pm$ 0.6\%} & -48.8\% & -2.2\% & \textbf{+3.1\%} & -10.1\% \\
        \midrule
        Pruning (0.2) & +15.5\% $\pm$ 6.0\% & +9.1\% $\pm$ 2.0\% & \underline{0.0\%} & -4.4\% & -41.0\% & -11.3\% \\
        Pruning (0.4) & +18.8\% $\pm$ 7.2\% & +6.2\% $\pm$ 1.7\% & \underline{0.0\%} & \underline{-7.3\%} & \underline{-61.0\%} & \underline{-20.6\%} \\
        Pruning (0.6) & \textbf{-67.9\% $\pm$ 1.4\%} & +7.3\% $\pm$ 1.6\% & \underline{0.0\%} & -5.9\% & -49.2\% & -18.2\% \\
        \midrule
        Quantization (float8) & +41.9\% $\pm$ 16.3\% & +98.3\% $\pm$ 3.2\% & -51.4\% & \textbf{0.0\%} & -1.1\% & \textbf{+0.4\%} \\
        Quantization (int8) & +102.2\% $\pm$ 14.4\% & +107.4\% $\pm$ 5.1\% & -51.4\% & -0.5\% & -0.9\% & -2.4\% \\
        Quantization (int4) & \underline{+133.5\% $\pm$ 26.3\%} & \underline{+201.6\% $\pm$ 4.8\%} & \textbf{-59.3\%} & -1.6\% & -4.4\% & -8.5\% \\
        \bottomrule
        \end{tabular}}
    \caption{RQ$_1$: Vulnerability Detection}
    \label{tab:defect_pred_time_size}
    \end{subtable}\\
% \vspace*{2mm}
    \begin{subtable}{\textwidth}
         \centering
            \resizebox{.8\textwidth}{!}{\begin{tabular}{l||c|c|c||c|c|c}
        \toprule
        \textbf{Compression} & \multicolumn{3}{c||}{\textbf{Efficiency}}  & \multicolumn{3}{c}{\textbf{Effectiveness}}\\
        \textbf{Method} & \textbf{CPU Inf. Time} & \textbf{GPU Inf. Time} & \textbf{Model Size}  & \textbf{Bleu} & \textbf{BERTScore} & \textbf{SIDE} \\
            \midrule
            \midrule
            None & 157.369 (sec) & 23.692 (sec) & 707 (MB) & 18.791 & 0.888 & 0.871 \\
            \midrule
            \midrule
            Know. Distil. & -39.8\% $\pm$ 7.8\% & \textbf{-2.2\% $\pm$ 4.9\%}$^*$ & -33.0\% & -42.3\% & -6.1\% & -70.6\% \\
            \midrule
            Pruning (0.2) & \textbf{-45.3\% $\pm$ 4.9\%} & +9.8\% $\pm$ 1.9\% & \underline{0.0\%} & -4.3\% & -0.1\% & \textbf{+0.4\%} \\
            Pruning (0.4) & +24.7\% $\pm$ 19.8\% & +121.9\% $\pm$ 13.1\% & \underline{0.0\%} & -66.6\% & -17.7\% & -42.4\% \\
            Pruning (0.6) & \underline{+183.5\% $\pm$ 41.9\%} & \underline{+419.3\% $\pm$ 29.5\%} & \underline{0.0\%}  & \underline{-93.4\%} & \underline{-58.4\%} & \underline{-93.0\%} \\
            \midrule
            Quantization (float8) & -20.3\% $\pm$ 7.9\% & +14.0\% $\pm$ 2.2\% & -42.0\% & \textbf{+0.4\%} & \textbf{0.0\%} & +0.1\% \\
            Quantization (int8) & -27.2\% $\pm$ 6.7\% & +6.2\% $\pm$ 2.3\% & -42.0\% & -0.3\% & \textbf{0.0\%} & +0.0\% \\
            Quantization (int4) & -17.9\% $\pm$ 7.0\% & +29.1\% $\pm$ 2.5\% & \textbf{-51.9\%} & -2.0\% & -0.1\% & +0.2\% \\
            \bottomrule
        \end{tabular}}
                \caption{RQ$_2$: Code Summarization}
        \label{tab:summ_time_size}
    \end{subtable}\\
% \vspace*{2mm}
    \begin{subtable}{\textwidth}
    \centering
    \resizebox{.8\textwidth}{!}{\begin{tabular}{l||c|c|c||c|c|c}
        \toprule
        \textbf{Compression} & \multicolumn{3}{c||}{\textbf{Efficiency}}  & \multicolumn{3}{c}{\textbf{Effectiveness}}\\
        \textbf{Method} & \textbf{CPU Inf. Time} & \textbf{GPU Inf. Time} & \textbf{Model Size}  & \textbf{MRR} & \textbf{MRR@1} & \textbf{MRR@5} \\
        \midrule
        \midrule
        None & 6.047 (sec) & 0.010 (sec) & 499 (MB) & 0.329 & 0.242 & 0.310 \\
        \midrule
        \midrule
        Know. Distil. & \textbf{-84.7\% $\pm$ 0.7\%} & \textbf{-29.2\% $\pm$ 0.5\%} & -48.7\% & -52.4\% & -57.7\% & -54.3\% \\
        \midrule
        Pruning (0.2) & +5.5\% $\pm$ 1.0\% & +11.1\% $\pm$ 0.7\% & \underline{0.0\%} & -3.2\% & -3.4\% & -3.4\% \\
        Pruning (0.4) & +16.1\% $\pm$ 2.2\% & +6.4\% $\pm$ 1.4\% & \underline{0.0\%} & -52.1\% & -57.3\% & -54.3\% \\
        Pruning (0.6) & +3.1\% $\pm$ 3.5\%$^*$ & +19.8\% $\pm$ 3.3\% & \underline{0.0\%} & \underline{-99.6\%} & \underline{-99.9\%} & \underline{-99.8\%} \\
        \midrule
        Quantization (float8) & +34.2\% $\pm$ 2.9\% & +113.2\% $\pm$ 1.8\% & -51.4\% & -0.2\% & \textbf{0.0\%} & -0.3\% \\
        Quantization (int8) & +42.2\% $\pm$ 3.8\% & +105.9\% $\pm$ 1.1\% & -51.4\% & \textbf{0.0\%} & -0.1\% & \textbf{+0.1\%} \\
        Quantization (int4) & \underline{+61.8\% $\pm$ 4.8\%} & \underline{+209.6\% $\pm$ 2.0\%} & \textbf{-59.3\%} & -6.3\% & -7.6\% & -6.7\% \\
        \bottomrule
    \end{tabular}}
        \caption{RQ$_3$: Code Search}
    \label{tab:code_search_time}
    \end{subtable}
\end{table*}

Table \ref{tab:results} shows the results for each RQ. On each sub-table, the first row reports the results of the plain CodeBERT model (i.e., without compression), while the remaining rows show the percentage variations provided by each compression strategy.\footnote{For inference time, the first row reports the median of measurements across batches of the plain CodeBERT model. The remaining rows show the percentage variations in the median inference time provided by each compression strategy, computed using the Kalibera and Jones approach \cite{Kalibera2012a}.} For inference time, we also report the confidence interval for the change using the Kalibera and Jones approach \cite{Kalibera2012a} (see Section \ref{sec:effic_metrics} for details). Non-statistically significant changes are marked with an asterisk ($^*$).
For each considered efficiency or effectiveness metric, the best values are highlighted in \textbf{bold}, while the worst values are \underline{underlined}.
\begin{figure*}[tb!]
\centering
\subfloat[RQ$_1$: Vulnerability Detection]{\includegraphics[width=0.9\linewidth]{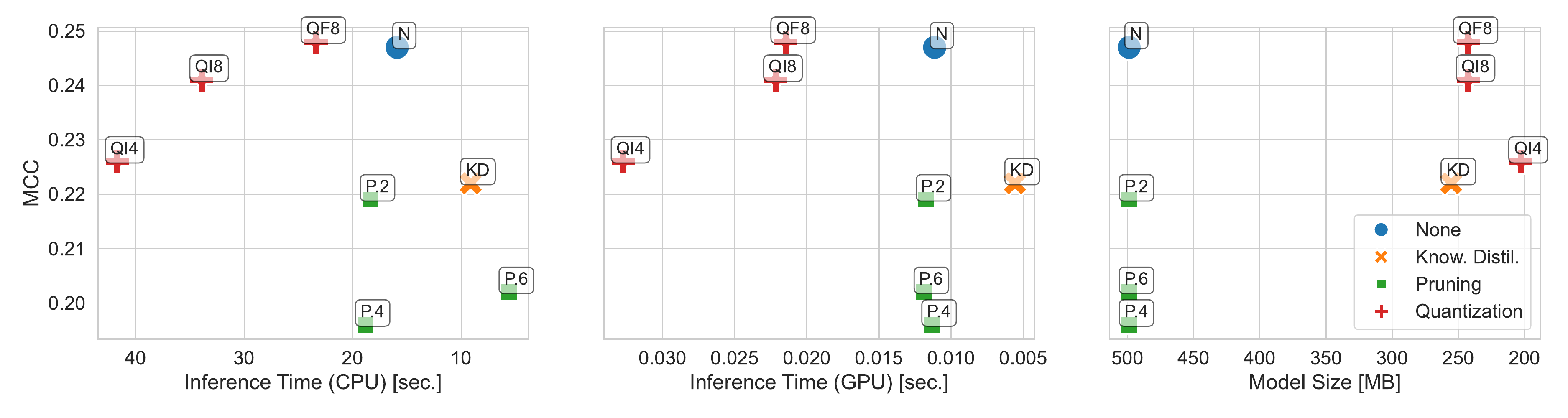}\label{fig:defect_pred_tradeoff}}\\
\subfloat[RQ$_2$: Code Summarization]{\includegraphics[width=0.9\linewidth]{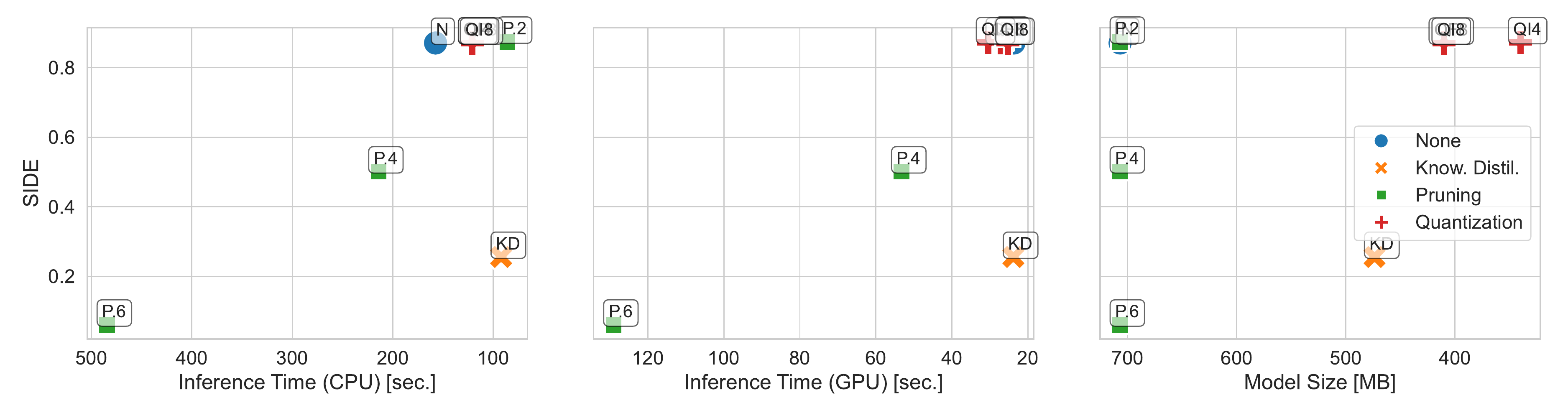}\label{fig:summ_tradeoff}}\\
\subfloat[RQ$_3$: Code Search]{\includegraphics[width=0.9\linewidth]{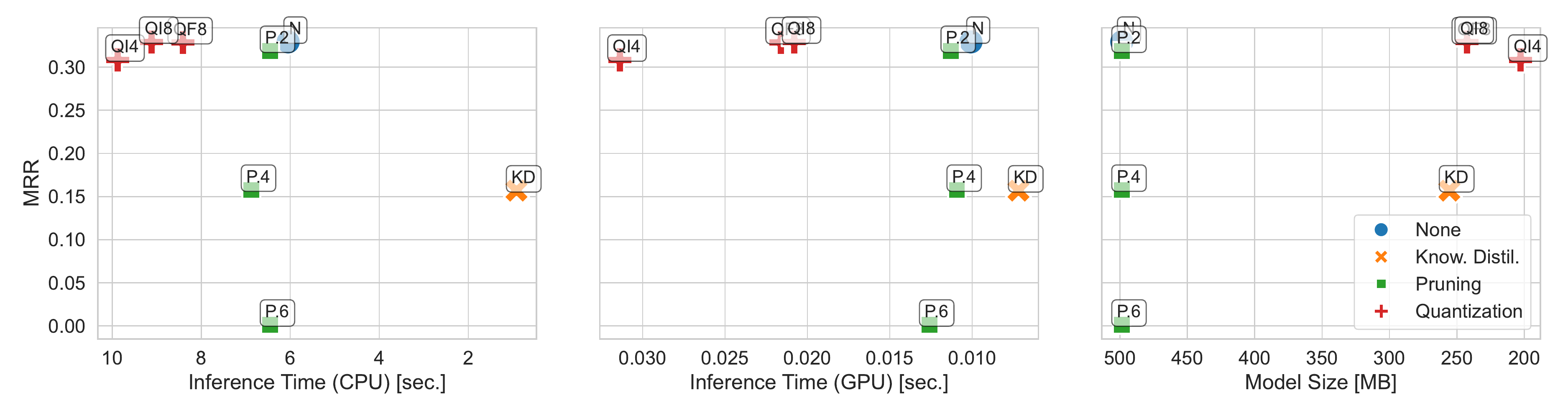}\label{fig:code_search_tradeoff}}
\caption{Trade-off between effectiveness (y-axis) and efficiency (x-axis) metrics for each of the SE tasks.}
\label{fig:tradeoff}
\end{figure*}
We also present scatter plots showing the trade-off between effectiveness ($y$-axis) and efficiency ($x$-axis) for each RQ in Figure \ref{fig:tradeoff}. 
%\subsection{RQ$_1$: \rqone}
\subsection{RQ$_1$ Results - Vulnerability Detection}
We analyse the impact of compression strategies from various aspects: inference time, model size, and vulnerability detection effectiveness. Additionally, we investigate the trade-offs involved among these metrics.
%This RQ focuses on the impact of compression strategies on an LM fine-tuned for the code vulnerability detection task.

% \input{tables/defect_pred_time_size}

\subsubsection{Inference Time}

%Table \ref{tab:defect_pred_time_size} reports the impact of compression strategies on the model's efficiency instead. 
As reported in Table \ref{tab:defect_pred_time_size}, all compression strategies change the inference time of the vulnerability detection model with statistical significance. However, the impact varies widely depending on the specific compression strategy and the hardware environment. On the CPU, for example, the most aggressive configuration of pruning (0.6) leads to the highest speed-up across all the strategies, with an inference time reduction of $-67.9\%$ compared to the plain CodeBERT model.
Interestingly, less aggressive forms of pruning (0.2 and 0.4) lead to the opposite outcome, with an inference slow-down of $+15.5\%$ and $+18.8\%$, respectively.
 All pruning configurations also negatively affect inference time in GPU environments, with slow-downs of up to $+9.1\%$.
A possible explanation for this behavior could be identified in the lower ability of GPU to handle sparse-matrix multiplications \cite{molchanov2017pruningconvolutionalneuralnetworks}.
These results of model pruning suggests that this strategy requires specific hardware and configurations to improve the inference time of vulnerability detection models; however, when these conditions are met, the benefits can be substantial.

From Table \ref{tab:defect_pred_time_size}, we also observe that quantization negatively impacts inference time in all configurations, resulting in the highest slowdowns across all compression strategies in both CPU and GPU environments. \texttt{int4} quantization, in particular, causes the most significant slowdowns, with increases of $+133.5\%$ on the CPU and $+201.6\%$ on the GPU.
%Moreover, we also notice how quantization strategies present a higher variability in the inference time compared with other compression models.
These negative results could be explained by the fact that CPUs and GPUs are heavily optimized for full-precision floating-point operations (e.g., 32-bit), hence quantized models may not fully take advantage of these optimizations and experience an inference slowdown \cite{gholami2022survey,nagel2020downadaptiveroundingposttraining,Shi2023,ganesh2021compressing}.

Knowledge Distillation, on the other hand, consistently and significantly improves inference time across both CPU and GPU environments. It ranks as the second-best strategy (after Pruning 0.6) on the CPU, with a reduction of $-39.8\%$ in inference time, and as the best strategy on the GPU, with a speed-up of $-47.7\%$.
%Quantization methods are instead confirmed to be the worst strategies, with \texttt{int4} quantization still being the worst among the three configurations ($+201.6\%$). Unlike CPU, Pruning 0.6 increases the model's inference time on GPU, with a percentage variation of $+7.3\%$. This behaviour aligns with the other pruning configurations, and a possible explanation could be identified in the lower ability of GPU to handle sparse-matrix multiplications \cite{molchanov2017pruningconvolutionalneuralnetworks}.

%\textcolor{blue}{BRAVI! Sezione molto ben scritta}

\subsubsection{Model Size}

As can be observed from Table \ref{tab:defect_pred_time_size}, all the compression strategies, with the exception of pruning, reduce the size the vulnerability detection model. 
\texttt{int4} quantization provides the highest model's size reduction ($-59.3\%$), followed by the other two quantization configurations ($-51.4\%$). Those results align with the expected quantization behaviour, where a lower bits' precision implies a lower model size.
Knowledge Distillation reduces the original model size by almost half ($-48.8\%$). This result aligns with the smaller architecture of the distilled model compared with the original fine-tuned CodeBERT model.
In contrast, pruning does not affect the model size. This behavior can be explained by the unstructured nature of the pruning strategy employed, where pruning impacts only the weight values by setting them to zero without modifying the network structure itself \cite{Shi2023}.

\subsubsection{Effectiveness}

% \input{tables/defect_pred_acc}

%Table \ref{tab:defect_pred_acc} 
%Table \ref{tab:defect_pred_time_size} reports the impact of compression strategies on the model's effectiveness for vulnerability detection.
Notably, we find that quantization has very limited impact on the model's effectiveness (see Table \ref{tab:defect_pred_time_size}).
\texttt{float8} quantization marginally changes the effectiveness, in terms of Accuracy, F1-score, and MCC ($0.0\%$, $-1.1\%$, and $+0.4\%$, respectively). Similar results hold for \texttt{int8} quantization, while \texttt{int4} quantization provides a slightly higher degradation, especially for F1 ($-4.4\%$) and MCC ($-8.5\%$).

Knowledge distillation performs slightly worse than quantization, particularly in terms of MCC ($-10.1\%$) and Accuracy ($-2.2\%$).
However, we observe an improvement in the F1-score ($+3.1\%$). This means that the distilled model has a higher tendency to predict vulnerabilities compared to the original fine-tuned CodeBERT model, which can decrease the number of true negative instances (i.e., increase the number of false positive instances). This leads to an improvement in recall, while only marginally affecting precision, which overall positively impacts the F1-score. Nonetheless, the reduction in MCC suggests a lower correlation between the predicted vulnerabilities and the actual ones. 

Pruning has the most significant detrimental impact on vulnerability detection effectiveness. All the pruning configurations considerably reduce model effectiveness, with MCC changes ranging from $-11.3\%$ to $-20.6\%$. Among all compression strategies, Pruning 0.4 performs the worst across all three metrics, showing effectiveness losses of $-7.3\%$, $-61\%$, and $-20.6\%$ for Accuracy, F1-score, and MCC, respectively. Interestingly, Pruning 0.6 performs slightly better than Pruning 0.4, despite prior research suggesting more severe effectiveness degradation with higher pruning levels \cite{gordon_compressing_2020}. %However, the difference in effectiveness between Pruning 0.4 and Pruning 0.6 is relatively small, especially in MCC, and could be attributed to the stochastic nature of the language model.
%\textcolor{blue}{questo spiega perchè avevamo un'efficienza maggiore...anche questo risultato è molto interessante!}
\subsubsection{Trade-offs}

% \begin{figure*}
%     \centering
%     \includegraphics[width=0.9\textwidth]{figures/defect_prediction_tradeoff.pdf}
%     \caption{Trade-off between MRR and efficiency metrics for vulnerability detection}
%     \label{fig:efect_pred_tradeoff}
% \end{figure*}

Figure \ref{fig:defect_pred_tradeoff} shows the effectiveness-efficiency trade-off provided by each compression strategy. For each sub-plot, the upper-right solutions are the best. We use MCC as the effectiveness metric since, as explained in Section \ref{sec:experiments}, it is the most comprehensive metric for classification \cite{MoussaDPmetrics}.

From the far-right subplot, we observe that quantization achieves the best model size reduction while maintaining effectiveness levels comparable to the baseline. However, as shown in the first two subplots, this gain comes at the cost of increased inference time on both CPU and GPU.
From Figure \ref{fig:defect_pred_tradeoff}, we observe that pruning is consistently outperformed by other strategies across all efficiency and effectiveness metrics. The only exception is Pruning 0.6, which offers the fastest inference time but at the expense of a comparatively low effectiveness.
Knowledge distillation is, on the other hand, the only compression strategy that improves efficiency across all dimensions (i.e., CPU and GPU inference time, and model size). At the same time, this strategy results in a comparatively moderate effectiveness degradation--greater than quantization but less than pruning. Overall, this makes knowledge distillation the strategy that offers the most balanced trade-off between efficiency gains and effectiveness degradation.
%The plots show that Knowledge Distillation, always in the middle-right portions of the plots, is the strategy that achieves the best effectiveness-efficiency trade-off concerning inference time and model size.
%This behaviour could be because pruning requires more in-depth knowledge of the network in selecting the layers and the amount of weights to prune. A wrong selection of those parameters may harm the overall model's behaviour \cite{gordon_compressing_2020}.

\begin{rqanswer}
    \textbf{Answer to RQ$_1$:} Quantization is the most effective strategy for reducing memory size in vulnerability detection models (up to $-59.3\%$), with minimal impact on model effectiveness ($-8.5\%$ MCC in the worst case). However, it can significantly increase inference time, by as much as $+201.6\%$.
    Pruning, on the other hand, shows no efficiency gains, with the exception of the 0.6 configuration that results in notable speed-up on CPU (up to $-67.9\%$), but at the cost of a $-18.2\%$ effectiveness degradation in MCC.
    Finally, Knowledge Distillation improves both inference time (up to $-47.7\%$) and model size (up to $-48.8\%$), while moderately impacting vulnerability detection effectiveness, with a reduction of $-10.1\%$ in MCC.
\end{rqanswer}

%\subsection{RQ$_2$: \rqtwo}
\subsection{RQ$_2$ Results - Code Summarization}
%We investigate the influence of various compression strategies on code summarization from various perspectives.

% \input{tables/summarization_time_size}

\subsubsection{Inference Time}
As shown in Table \ref{tab:summ_time_size}, Pruning 0.2 is the most effective approach for reducing inference time on the CPU, achieving a $-45.3\%$ reduction compared to the plain CodeBERT model. Counter intuitively, we observe a negative correlation between the percentage of weights pruned and the inference time reduction. For example, Pruning 0.4 performs worse than Pruning 0.2, increasing inference time by $+24.7\%$, while Pruning 0.6 shows the worst  behavior, increasing inference time by $+183.5\%$ on the CPU.
On the GPU, we observe an even more pronounced worsening trend, as inference times increase by $+9.8\%$, $+121\%$, and $+419.3\%$ for Pruning 0.2, 0.4, and 0.6, respectively.
%In addition, we observe how Pruning 0.6 provides a higher variability in the model's inference time.
A possible explanation for this behaviour could be the lower hardware's ability to handle sparse matrix multiplications \cite{liu2019rethinkingvaluenetworkpruning}. Since generation tasks require multiple network forward steps, a higher sparsity of the weights could increase the overall inference time.

Quantization strategies can all reduce the inference time on the CPU, with \texttt{int8} quantization being the best configuration ($-27.2\%$). This behaviour differs from what we observed for vulnerability detection and could be explained by the higher complexity of the underlying task, which may benefit more from reduced weight precision \cite{gholami2022survey}.
On the other hand, on the GPU, quantized models are overall slower than the plain CodeBERT model, with inference time slow-downs ranging from $+6.2\%$ to $+29.1\%$.

Knowledge Distillation is the only strategy capable of reducing inference time of code summarization models across both CPU and GPU environments. On the CPU, it is the second-best compression strategy for inference time reduction  ($-39.8\%$). On the GPU, it is the only strategy to achieve an inference time reduction, with a decrease of ($-2.2\%$). However, this reduction is not statistically significant, as the confidence interval for the relative change includes zero (see Section \ref{sec:effic_metrics} for details).

\subsubsection{Model Size}

From Table \ref{tab:summ_time_size}, we observe how, like in vulnerability detection, \texttt{int4} quantization is the best strategy for reducing the model's size ($-51.9\%$), followed by the other two quantization configurations. Again, these results align with the expected quantization behaviour, where the lower the precision, the lower the size. Knowledge Distillation can also reduce the model size (by $-33\%$), while the adopted pruning strategies confirms that they do not lead any change.   

\subsubsection{Effectiveness}

From Table \ref{tab:code_search_time} we observe that all quantization strategies provide only marginal change in the model's effectiveness, with \texttt{float8} and \texttt{int8} quantization having almost comparable metrics to the baseline, and \texttt{int4} quantization showing limited degradations of $-2\%$, $-0.1\%$, $-0.2\%$ for Bleu, BERTScore and SIDE, respectively.

For pruning, we observe that the impact grows as pruning becomes more aggressive. Specifically, Pruning 0.2 shows marginal impact, with a slight degradation of $-4.3\%$ in summary-summary text similarity (BLEU). On the other hand, Pruning 0.4 and 0.6 substantially reduce the model's effectiveness across all dimensions. For instance, in terms of summary-code semantic similarity (SIDE), Pruning 0.4 and 0.6 lead to effectiveness decreases of $-42.4\%$ and $-93\%$, respectively.
This behaviour is in line with previous research showing how the effectiveness of a BERT model starts to decrease if the amount of pruned weights is $\geq 40\%$ \cite{gordon_compressing_2020}.

Knowledge Distillation also shows significant impact, particularly in terms of summary-summary text similarity ($-42.3\%$ in Bleu) and summary-code semantic similarity ($-70.6\%$ in SIDE).
This behavior is consistent with previous studies, which highlight that distilled models are less effective for tasks beyond classification \cite{sanh2020distilbertdistilledversionbert}.

\subsubsection{Trade-offs}

% \begin{figure*}[tb]
%     \centering
%     \includegraphics[width=.9\textwidth]{figures/summarization_tradeoff_median.pdf}
%     \caption{Trade-off between SIDE and efficiency metrics for Code Summarization}
%     \label{fig:summ_tradeoff}
% \end{figure*}

Figure \ref{fig:summ_tradeoff} illustrates the trade-off between effectiveness and efficiency. We use SIDE as the effectiveness metric, as it is specifically designed for code summarization tasks \cite{mastropaolo_evaluating_2023}.
We observe that Knowledge Distillation overall improves the inference time and size of the model, but it significantly degrades effectiveness.
In contrast, quantization generally performs well across all efficiency metrics, with only a slight slow-down in inference time on the GPU. Moreover, this strategy has only a marginal impact on the model's effectiveness, with values comparable with the baseline.
As such, quantization appears to offer the best balance between efficiency and effectiveness as a compression strategy.
Pruned models are generally outperformed by other compressed models in terms of both efficiency and effectiveness. The only exception is Pruning 0.2, which provides the best trade-off between inference time and effectiveness on the CPU. However, using pruning does not improve model size.

%we observe an almost inverse linear relationship between the percentage of weights pruned and pruning performances regarding inference time and effectiveness.

\begin{rqanswer}
    \textbf{Answer to RQ$_2$:} Quantization strategies offer the best efficiency-effectiveness trade-off among compression techniques, with inference time speed-up of up to $-27.2\%$, model size reduction of up to $-51.9\%$, and minimal effectiveness degradation of up to $-0.2\%$ in SIDE. Pruning generally underperforms compared to other strategies, but under specific configurations, such as Pruning 0.2, it achieves the best inference time improvements on the CPU ($-45.3\%$). While Knowledge Distillation improves all efficiency metrics (with up to $-39.8\%$ in inference time and $-33\%$ in model size), it causes significant losses in effectiveness, with reductions of $-70.6\%$ in SIDE.
\end{rqanswer}

%\subsection{RQ$_3$: \rqthree}
\subsection{RQ$_3$ Results - Code Search}

\subsubsection{Inference Time}

Table \ref{tab:code_search_time} reports how Knowledge Distillation is the strategy that better reduces the inference time on both CPU and GPU, with improvements of $-84.7\%$ and $-29.2\%$, respectively.
We observe that all other compression strategies have a negative impact on inference time. Pruning slows down inference on both CPU and GPU, with an increase in inference time ranging from $+3.1\%$ to $+19.8\%$. A possible explanation for this behavior with pruning could be the complexity of the code search task, which typically requires multiple comparisons between a code comment and the code snippets. In that, the sparsity of matrices could increase inference time, especially if the hardware is not well-optimized for handling sparse matrices \cite{molchanov2017pruningconvolutionalneuralnetworks}.
Similarly to vulnerability detection, we observe that quantization consistently increases the inference time of code search on both CPU and GPU. The magnitude of the increase is lower on CPU, ranging from $+34.2\%$ to $+61.8\%$, and higher on GPU, with variations from $+113.2\%$ to $+209.6\%$. The negative behavior of quantization strategies could be (once again) explained by the lack of optimization in CPU and GPU kernels for low-precision bit operations \cite{gholami2022survey,nagel2020downadaptiveroundingposttraining,Shi2023,ganesh2021compressing}. 
\subsubsection{Model Size}

Since the model used for this task is the same as the one used for vulnerability detection (i.e., CodeBERT), the results regarding the model size are identical. Hence, \texttt{int4} quantization emerges as the best strategy, while the unstructured nature of pruning does not imply any change. 

\subsubsection{Effectiveness}

Table \ref{tab:code_search_time} shows how \texttt{int8} and \texttt{float8} quantization provide no significant change compared with the baseline. A slightly higher degradation is instead observed with \texttt{int4} quantization ($-6.3\%$ in MRR).
All pruning strategies provide a degradation in the model's effectiveness, with reductions in MRR ranging from $-3.2\%$ to $-99.6\%$. We observe a correlation between the increase in effectiveness loss and the percentage of pruned weights.
 Pruning 0.6 strategy proves to be the worst compression strategy, with an MRR loss of $-99.6\%$.
%This behaviour again aligns with what has been observed in previous research \cite{gordon_compressing_2020}.
Finally, we also observe a significant degradation in effectiveness for Knowledge Distillation, with a $-52.1\%$ loss in MRR. This result is consistent with previous research, which has shown that distilled models often exhibit lower effectiveness on tasks different from classification \cite{sanh2020distilbert}.

\subsubsection{Trade-offs}

% \begin{figure*}[tb]
%     \centering
%     \includegraphics[width=.9\textwidth]{figures/code_search_tradeoff_median.pdf}
%     \caption{Trade-off between MRR and efficiency metrics for Code Search}
%     \label{fig:code_search_tradeoff}
% \end{figure*}

Figure \ref{fig:code_search_tradeoff} shows the effectiveness-efficiency trade-off. We adopt the general MRR score as the effectiveness metric.
We observe that quantization strategies are preferred to reduce the model size while not impacting its effectiveness; however, they negatively influence the inference time.
%Comparing the quantization strategies, \texttt{int4} quantization better reduces the model size but increases more the inference time, while the opposite holds for \texttt{int8} and \texttt{float8} quantization.
Knowledge Distillation achieves significant reductions in both inference time and model size, but it also drastically reduces the model’s effectiveness, with a loss of $-52.1\%$ in MRR.  Finally, pruning strategies do not achieve positive results in any of the efficiency and effectiveness metrics analysed.

\begin{rqanswer}
    \textbf{Answer to RQ$_3$:} Quantization strategies drastically reduce the size of code search models (up to $-59.3\%$) with only a marginal impact on effectiveness (up to $-6.3\%$ in MRR). However, they can significantly slow down inference time, with an increase of up to $+61.8\%$ on CPU and $+209.6\%$ on GPU. Knowledge Distillation reduces both model size ($-48.7\%$) and inference time ($-84.7\%$ on CPU and $-29.2\%$ on GPU), but it comes at the cost of a considerable loss in effectiveness ($-52.4\%$ in MRR). Pruning, however, does not show improvements in any of the efficiency metrics analysed. 
\end{rqanswer}

\section{Discussion}\label{sec:discussion}
In the following, we discuss the overall behaviour of the analysed compression strategies and report insights for practitioners and researchers derived from our empirical evaluation. 

\subsection{Performance of LM Compression Strategies}

\subsubsection{Knowledge Distillation} We found that Knowledge Distillation is the only compression strategy capable of improving both inference time and model size across all the software engineering tasks we considered. However, this strategy can lead to a significant loss in effectiveness. This loss in effectiveness is particularly pronounced in tasks like code search and summarization, while it is milder in vulnerability detection. These findings align with previous research, which suggests that distilled models tend to be less effective for tasks other than classification \cite{sanh2020distilbert}. Indeed, to date, knowledge distillation has predominantly been applied to code classification tasks, such as vulnerability detection and clone detection \cite{Shi2023, Shi2024}.

\subsubsection{Model Quantization} We found that quantization drastically reduces the size of models across all tasks while maintaining relatively high effectiveness. However, this improvement often comes at the cost of increased inference time.
On GPU environments, quantization can double or even triple the inference time for tasks like vulnerability detection and code search. In contrast, the slowdown in inference time is much less pronounced for the code summarization task.
Interestingly, in CPU environments, quantization can even improve the inference time for code summarization. However, for vulnerability detection and code search, it still introduces notable slowdowns.
Based on these findings, we hypothesize that quantization performs better in terms of efficiency when the task is complex—requiring multiple forward passes through the model—such as in code generation tasks. These results align with previous studies on the application of quantization for code generation tasks \cite{Wei2023}.

%In particular, we observed that the CPU's inference time significantly increases if the inference time of the non-compressed model is not high (like in vulnerability detection or code search tasks) and that this pattern is even amplified on the GPU, where the latency time is never improved. This behaviour may be explained by a lower optimization of kernels (especially GPUs) for low-precision bit operations, which may increase the overall latency  \cite{gholami2022survey,nagel2020downadaptiveroundingposttraining,Shi2023,ganesh2021compressing}. 
%Regarding quantization precision, we found that using \texttt{int4} leads to better model size reduction but impacts the inference time more. On the other hand, \texttt{float8} and \texttt{int8} quantization only differ in defect prediction efficiency, with \texttt{int8} resulting in higher inference time.  %In terms of effectiveness degradation, the only observed difference is in defect prediction, where lower precision leads to lower effectiveness. However, the difference in effectiveness between non-compressed and quantized models is always negligible.

\subsubsection{Model Pruning} We found that while pruning offers limited overall benefits in terms of efficiency, it can lead to significant speed-ups in inference time under specific combinations of configurations and environments. For instance, Pruning 0.2 provides the highest inference speed-up for code summarization on CPU among all compression strategies, with only a marginal reduction in effectiveness. Similarly, in the vulnerability detection task, Pruning 0.6 proves to be the most effective strategy for improving inference time on CPU, though it comes with a moderate loss in effectiveness. These findings suggest that, when properly configured, pruning can deliver substantial inference speed-ups, particularly in CPU environments.

%Concerning inference time, pruning can never reduce the GPU latency time. However, on CPU we observe different behaviours based on the specific task analysed. Pruning 0.6 can significantly reduce the inference time for defect prediction while Pruning 0.2 is more effective on code summarization. The variability in the tasks analyzed could potentially account for this behaviour. When pruning strategies are employed, the sparsity of the weight matrix in a network increases. However, hardware, particularly GPUs, may not be as optimized for sparse matrix multiplication. Since generative (like code summarization) and search (like code search) tasks involve multiple forward passes in a network, an increase in sparsity could potentially lead to longer inference times \cite{liu2019rethinkingvaluenetworkpruning}. Finally, as confirmed by previous research \cite{gordon_compressing_2020}, a percentage of pruned weights $\geq 40\%$ implies a higher model's effectiveness degradation. 

\subsection{Insights}

\subsubsection{Insights for Practitioners}

Our results indicate that the impact of different compression strategies can vary significantly depending on the SE task and the underlying execution environment, often involving important efficiency-effectiveness trade-offs. Hence, practitioners should carefully select the compression strategy based on their requirements and the underlying task as follows:

\begin{itemize}[leftmargin=0.45cm]
    \item  If the practitioner's priority is to improve both inference time and model size, regardless of the task or underlying execution environment, Knowledge Distillation is the preferred choice. Indeed, it is the only compression strategy that delivers improvements across all efficiency aspects in both CPU and GPU environments.  However, practitioners should be aware of potential negative impacts on effectiveness, particularly in tasks other than code classification (e.g., code summarization and code search). Additionally, it is important to note that Knowledge Distillation is the only compression strategy that requires re-training a model from scratch, which may be undesirable if practitioners lack sufficient computational resources.

   \item If the priority is to reduce model size without significantly degrading effectiveness, quantization is the natural choice, regardless of the SE task. Nevertheless, practitioners should be mindful of the potential side effects on inference time, which can vary greatly depending on the task and environment, and often result in significant slowdowns. In some specific cases, however, quantization can even improve inference time, such as code summarization on CPU.

   \item If the practitioner's goal is to reduce inference time in GPU environments, the only viable choice is Knowledge Distillation. However, as previously mentioned, this approach can drastically affect model effectiveness, particularly for tasks like code summarization and code search.
   
   \item If the priority is to improve inference time on CPU, the practitioner could once again consider Knowledge Distillation. They may also consider pruning, especially for tasks such as vulnerability detection and code summarization. However, pruning must be carefully configured to provide benefits. For instance, a less aggressive form of pruning (0.2) proved beneficial in reducing inference time for vulnerability detection, while a more aggressive form (0.6) resulted in the highest inference speed-up for code summarization. For code search, Knowledge Distillation remains the only viable option for reducing inference time on CPU, though it drastically reduces the model's effectiveness. In fact, we did not find any compression strategy that improves inference time for code search without significant losses in effectiveness.

    %\item If the priority is on inference time reduction, then Knowledge Distillation is the only strategy effective on both CPU and GPU. However, it may cause a significant loss in effectiveness if the underlying task differs from code classification. In addition, it is worth mentioning that Knowledge Distillation is the only compression strategy that requires fine-tuning a model from scratch, which may create issues if the practitioner lacks the computational resources to fine-tune a model.
    %\item Quantization and Pruning(0.2) are instead effective in inference time reduction only on CPU and for more computationally intensive tasks like code summarization. Finally, we discourage pruning the model's weights for an amount $\geq 40\%$, since it may cause significant degradation in effectiveness. 

    %\item If the task at hand cannot allow for any loos in effectiveness, no compressed models should be applied. \todo{Giordano pls double-check my claim}
    
\end{itemize}

\subsubsection{Insights for Researchers}

Our results show that the behaviour of compression strategies greatly varies depending on the context.
However, when these strategies are carefully selected for the underlying task and environment, they can significantly enhance efficiency aspects with a marginal impact on effectiveness. For instance, we found that specific quantization configurations can enhance both CPU inference time and model size without influencing the effectiveness of code summarization. These findings highlight the potential for developing approaches that automatically select the optimal compression strategy based on the underlying task, execution environment, and efficiency/effectiveness requirements. Moreover, we observed that the efficacy of compression strategies can highly depend on their specific configuration. This behavior is particularly evident in the pruning strategy, where the amount of pruned weights significantly influence its impact on inference time. We encourage future research aimed at developing approaches to automatically identify the optimal compression configuration (e.g., amount of weights to prune), based on the SE task and execution environment.

%For instance, one possible avenue would be to focus on automatically identifying the proper parameters and amount of weights to prune in a model based on the specific task. %Researchers could also focus on improving quantization performances in cases where the inference time of the non-compressed baseline is not high.

%No compression strategies achieve an optimal behaviour with respect to both effectiveness and efficiency. %and each compression strategy presents some trade-offs regarding effectiveness and efficiency gain. Hence, future research should address these challenges from both hardware and software perspectives.Possible research concerning hardware could focus on improving CPU and GPU ability to handle low-precision \cite{gholami2022survey,ganesh2021compressing} and sparse matrix operations \cite{liu2019rethinkingvaluenetworkpruning}. Possible research concerning software should focus on designing new compressing strategies or improving existing ones to reduce both inference time and model size while maintaining high levels of effectiveness. One possible avenue would be for researchers to focus on automatically identifying the proper parameters and amount of weights to prune in a model based on the specific task. Researchers could also focus on improving quantization performances in cases where the inference time of the non-compressed baseline is not high. 

\section{Threats to Validity}\label{sec:threats}
\textit{Internal Validity:} Execution time measurements are typically subject to variability, which can hinder rigorous evaluation \cite{Maricq2018}. To address this issue, we followed best practices from performance engineering to increase the reliability of our evaluation \cite{Kalibera2012a, Kalibera2013a,Traini2024, Jangali2023a, Zhang2023a, Laaber2020a} (see Section \ref{sec:effic_metrics}). Furthermore, the outcomes may be influenced by potential errors in the implementation of the baseline models and compression strategies. To address this concern, we utilized a widely adopted benchmark (CodeXGLUE) for training and testing the baseline models and employed compression strategies implementations from widely used and maintained libraries. 

\textit{Construct Validity:} As explained in Section \ref{sec:effec_metrics}, we employed multiple metrics for each task to assess a model's effectiveness. This has been done to account for possible threats concerning adopting specific metrics (like Accuracy \cite{MoussaDPmetrics} or Bleu \cite{sun_source_2024}). Concerning efficiency metrics, we relied on standard approaches and statistics to measure inference time and model size.

%\textit{Conclusion Validity:} Some results concerning inference time are not statistically significant. This could be explained by the variability in the measurements over multiple batches. However, this does not influence the overall conclusions drawn from our evaluation. 

\textit{External Validity:} The major threats of our work concern its generalizability. The results obtained are limited to the models and tasks analysed and the environment in which the experiments were run. In addition, the results concerning Knowledge Distillation are specific to the DistilBERT LM and may not hold for other distilled models. Future work can extend our analysis with other LMs and code-related tasks and by analysing other Knowledge Distillation techniques. To this end we strove to describe the methodology we followed as clearly as possible and have made our code and scripts publicly available  \cite{repl_package}.

\section{Conclusion and Future Work}\label{sec:concl}
In this paper, we performed an empirical evaluation of the impact that using three LM compression strategies (i.e., Knowledge Distillation, Quantization, and Pruning) could have on the inference time, memory size, and effectiveness of models fine-tuned for three widely adopted SE tasks -- vulnerability prediction, code summarization, and code search. 
Results show how each strategy provides some trade-offs among the three analysed dimensions and how the proper compression method should be chosen based on the underlying task and practitioner's priorities.
Future work includes extending our analysis to other SE tasks, language models (like CodeT5 or GraphCodeBERT), compression strategies and energy metrics. 
%One could compare the impact of using compression strategies on other LM for code (like CodeT5 or GraphCodeBERT), as well as analyse other compression strategy implementations, like the ones proposed by Shi et al. \cite{Shi2023,Shi2024}. 
Another interesting avenue would be the analysis of how compression strategies perform in devices with very limited computational resources. %Finally, our analysis could be extended by including other metrics related to energy consumption.

\footnotesize
\section*{Acknowledgment}
This work is supported by European Union - NextGenerationEU - National Recovery and Resilience Plan (Piano Nazionale di Ripresa e Resilienza, PNRR) - National Centre for HPC, Big Data and Quantum Computing (CN\_00000013  – CUP: E13C22001000006) and by \EmeliotAck and by Territori Aperti, a project funded by Fondo Territori Lavoro e Conoscenza CGIL CISL UIL.
\HPCAck

\normalsize

\bibliographystyle{IEEEtran}
\bibliography{bibliography}

\end{document}